# The calculation of the diffusion coefficients in ternary multiphase Ti-NiAl system

Daria Serafin, Wojciech J. Nowak, Bartek Wierzba

Department of Materials Science, Faculty of Mechanical Engineering and Aeronautics, Rzeszow University of Technology, al. Powstańców Warszawy 12, 35-959 Rzeszów, Poland

corresponding author: Bartek Wierzba - bwierzba@prz.edu.pl

**Abstract.** Determination of diffusion coefficients in a ternary multiphase system is discussed as the data on the approximation method allowing to determine the diffusion coefficients in such systems are currently missing in the literature. In the present work, the mass conservation approach (MCA) for calculation the diffusivities in individual phases with the linear differences in composition is proposed. The diffusion data can be determined exclusively based on the experimental results. The method is implemented and validated for the ternary Ti-NiAl system, in which three intermetallic phases are stable.

**1. Introduction.** The most important parameter describing the diffusion process is diffusion coefficient, $D_i$. It was historically introduced by Fick′s in his famous diffusion laws. The first Fick′s law describes the process with constant mass in time - the flux, Eq.1.

$$j_i = -D_i \frac{\partial N_i}{\partial x} \qquad (1)$$

where: $j_i$ is the diffusion flux and $N_i$ molar ratio of the i-th component.

The second Fick′s law describes how the mass changes with time during the diffusion process, Eq. 2.

$$\frac{\partial N_i}{\partial t} + \frac{\partial j_i}{\partial x} = 0 \qquad (2)$$

Diffusion coefficients help describe the diffusion related phenomena such as mechanical properties of materials, i.e. stress induced diffusion [1], phase transformations [2], etc. In many contact systems such as coatings, diffusion is the rate limiting step controlling the growth of intermetallic compounds. The diffusion coefficients can be determined experimentally using the tracer and diffusion coupling method [3]. However, these methods are seldom applicable to multicomponent systems [4]. A more straightforward way to determine diffusivities in such systems is to use these techniques for binary systems. The methods in question have been modified by introducing additional parameters like the position of the Kirkendall plane so that only one experiment is required for determination of diffusion coefficient [4, 5, 6, 7, 8, 9, 10, 11, 12, 13]. Ternary and higher order systems are substantially more complicated for experimental studies. The main and cross diffusion coefficients can be determined at the point of intersection of the composition profiles [14]. The biggest disadvantages of the above mentioned methods is their inapplicability to multiphase systems. A reasonable method to determine diffusion coefficient in multicomponent and multiphase systems is currently missing. Furthermore, the traditional experimental techniques are extremely expensive and time consuming.

With the advent of powerful computers, the diffusion coefficients are now easily calculated using numerical algorithms. The computational methods are more effeciant compared to the traditional ones. Moreover, the compositional dependence of diffusion coefficients should be taken into account to increase the accuracy. [15, 13]. Therefore, the demand for novel efficient techniques providing reliable diffusion coefficients in multicomponent alloys is constantly growing.

The compositional dependency of interdiffusion coefficients in binary and ternary alloys has been extensively studied by Fujita and Gosting [16] and later by other authors [17, 18, 19, 20, 21, 22]. Cermak and Rothova [23] computed the composition dependent diffusion coefficient deviding the range into small volume cells and integrating the diffusion equations. based on the Fick's second law, Bouchet and Mevrel [19] proposed an inverse method to calculate the composition dependent intrinsic diffusion coefficients. All methods above are based on the mass conservation law and are thus deterministic ones. However, stochastic methods based on the genetic algorithms or monte carlo simulations are also found in literature in application to the composition dependent diffusion coefficient problem. These methods are superior to the deterministic ones and can be used for calculating diffusion coefficients in a wide concentration range.

The situation dramatically changes when diffusion coefficients should be evaluated in multiphase systems. Literature is very limited for such cases. One of the classical methods to calculate the diffusion coefficient in a binary multiphase system was proposed by Wagner [24, 25]. It allowed calculating the average intrinsic diffusion coefficient over the entire range. The integral diffusion coefficient can be calculated from the following expression:

$$\tilde{D}(N_A^*) = \frac{(N_A^+ - N_A^-)V_B^*}{2t(\partial N_A/\partial x)_{x=x^*}} \left[ (1-Y^*)\int_{-\infty}^{x^*} \frac{Y}{V} dx + Y^* \int_{x^*}^{\infty} \frac{1-Y}{V} dx \right] \quad (3)$$

where: $N_A^-$ and $N_A^+$ are the molar ratio of A and B component at the boundary of the system, $V = 1/C$ is the volume of the system. The variable Y can be expressed as:

$$Y = \frac{N_A - N_A^-}{N_A^+ - N_A^-} \qquad (4)$$

The Wagner's integral coefficient can be used in binary multiphase systems.

In the literature, the methods calculate diffusion coefficients in multicomponent (even ternary) multiphase systems are rather scarce. Thus, in the present work paper the deterministic method of approximating intrinsic diffusion coefficients based on the Leibnitz rule at the phase boundaries is presented. The obtained intrinsic diffusion coefficients are valid for multiphase systems. Finally, the novel procedure was verified for a ternary Ti-NiAl system.

## 2. Modeling
### 2.1. Pseudobinary approach (PBA) in a ternary system

As mentioned above, the methods to calculate diffusion coefficients in ternary systems are scarce. However, a few approaches have been tested in order to estimate the coefficients assuming a number of simplifications. Based on Wagner's equations [4], Paul proposed a method of calculating diffusion coefficient considering a ternary A-B-C system and a diffusion couple, in which component C does not diffuse substituting component B in the alloy. The concentration of C remains constant during the diffusion process (table 1). Paul assumed further that a new phase γ grows in the middle of initial blocks, the composition changing linearly in the interdiffusion zone within a range of $\Delta N_A^\gamma = 0.2 - 0.3$. This variation does not affect the entire molar volume (Fig. 1). Thus, Eq. 3 for calculation the interdiffusion coefficient may be reformulated as follows:

$$\tilde{D} = \frac{1}{2t}\left(\frac{dx}{dY_A}\right)_K \left[(1-Y_A^K)\int_{-\infty}^{x_K} Y_A dx + Y_A^K \int_{x^*}^{\infty}(1-Y_A)dx\right] \qquad (5)$$

where $K$ is the marker plane position. Considering the diffusion time of 25 hours and thickness of γ phase layer as 90 µm, the marker plane had been found at 40.5 µm from Alloy 1– γ phase interface of the diffusion couple (Fig. 2). The concentration of component A at the marker plane is 24,5 at. %.

Table 1. Initial composition of the components

|  | A concentration, at. % | B concentration, at. % | C concentration, at. % |
|---|---|---|---|
| Alloy 1 | 15 | 75 | 10 |
| Alloy 2 | 35 | 55 | 10 |

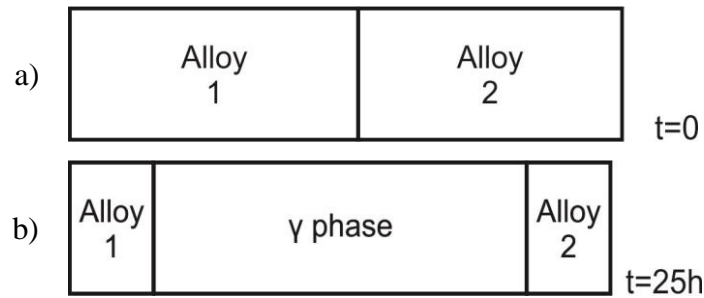

Fig. 1. Diffusion couple of two alloys: a) before diffusion process, b) after annealing for 25h

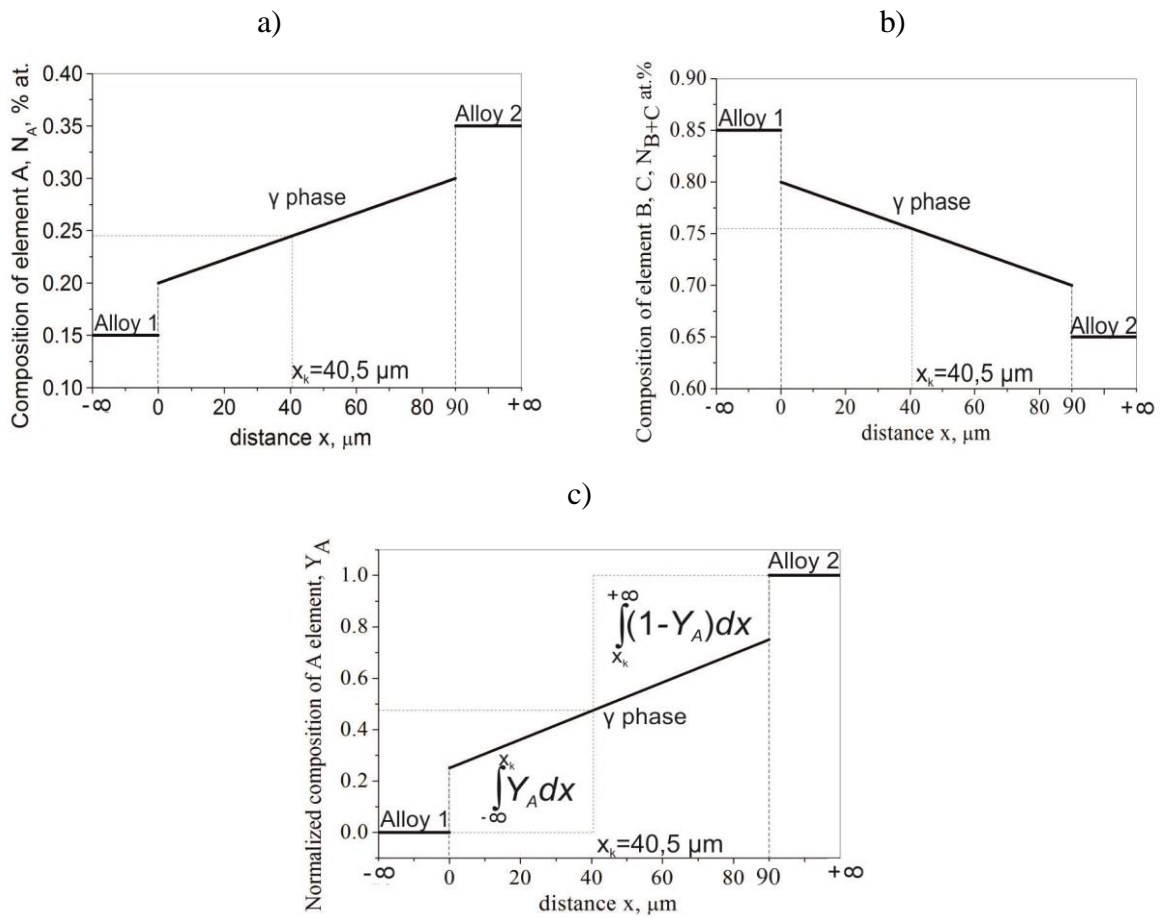

Fig. 2. Composition N of elements: a) A, b) B and C, c) normalized composition of A element $Y_A$

Paul assumed that component B substitutes component C so that the concentration profile of component A is not affected by the addition of component C to the alloy. When the interdiffusion coefficient are calculated (Eq. 5), the concentrations of components B and C have no effect on the outcome, i.e. the compositional range of the diffusion couple is normalized (Eq. 4). To calculate the of intrinsic diffusion coefficients, however, the concentration profiles of components B and C are crucial if the relationship between the interdiffusion and intrinsic diffusion coefficients is considered:

$$\tilde{D}^K = N_A D_B^K + N_B D_A^K \qquad (6)$$

Should the concentration of component C be neglected, the conservation equation for the ternary system: $N_A + N_B + N_C = 1$ is not valid. To avoid this problem in calculation of intrinsic diffusion coefficients, the following concentration profiles have been used: $N_A$ and $N_{B+C}$ (or in a normalized form: $Y_A$ and $Y_{B+C}$) in the displacement function (Fig. 2b). Thus, the values of $D_A$ and $D_B$ are calculated as follows:

$$D_A = \frac{1}{2t}\left(\frac{dx}{dN_A}\right)_K \left[ N_A^+ \int_{x^{-\infty}}^{x_K} Y_A dx - N_A^- \int_{x_K}^{x^{+\infty}} (1-Y_A) dx \right] \qquad (7)$$

$$D_B = -\frac{1}{2t}\left(\frac{dx}{dN_A}\right)_K \left[ N_{B+C}^+ \int_{x^{-\infty}}^{x_K} Y_A dx - N_{B+C}^- \int_{x_K}^{x^{+\infty}} (1-Y_A) dx \right] \qquad (8)$$

Combining the assumptions made by Paul with the equations above, the following values for interdiffusion and intrinsic diffusion coefficients were calculated: $\tilde{D} = 1.68 \cdot 10^{-14}$, $D_A = 1.13 \cdot 10^{-14}$, $D_B = 3.38 \cdot 10^{-14}\,\mathrm{m^2/s}$.

## 2.2. Interdiffusion in forming intermetallic compounds by average atom movement (AAM)

In the Paul's example, all the experimental data is known, i.e. concentration of all elements, width of the diffusion region, time, and position of marker plane. Thus, the interdiffusion coefficient of γ phase may be calculated using simple diffusion equation:

$$(\Delta x)^2 = 4\tilde{D}t \qquad (9)$$

Transforming equation (Eq. (9)) and substituting the data from section 2.1, one to obtains the value of the interdiffusion coefficient in γ phase: $\tilde{D} = 2.25 \cdot 10^{-14}\,\mathrm{m^2/s}$.

Assuming further that the atomic volume does not vary across the region, the following relation can be derived:

$$\frac{J_B}{\beta J_A} = \frac{x_K - x_L}{x_R - x_K} \tag{10}$$

where fluxes $J_A$ and $J_B$ are calculated from the Fick's law (Eq. 11) and $\beta$ is a stoichiometric subscript indicating the growing phase ($A_\beta B$, $\beta=1,2,\ldots$). Combining Eqs (1) and (10) for both elements A and B, the following expression is obtained:

$$\frac{D_B}{D_A} = \frac{\beta(x_K - x_L)}{x_R - x_K} \tag{11}$$

Knowing the value of the interdiffusion coefficient, the assumption that $\beta = 2$ in connection with (Eq. (6)) and (Eq. (11)) makes it possible to calculate the intrinsic diffusion coefficient for elements A and B using the following equations:

$$D_B = \frac{\tilde{D}}{N_A + N_B \dfrac{x_R - x_K}{\beta(x_K - x_L)}}$$

$$D_A = \frac{D_B(x_R - x_K)}{\beta(x_K - x_L)} \tag{12}$$

The obtained values are $D_A = 1.95 \cdot 10^{-14}$ and $D_B = 3.19 \cdot 10^{-14}$ m$^2$/s, respectively.

## 2.3. Mass conservation law (MCA)

Interdiffusion and intrinsic diffusion coefficients can be also calculated taking into account the mass conservation law applying to diffusion processes, interdiffusion between two alloys. If only one intermetallic phase $\gamma$ grows in the diffusion couple, diffusion can be decribed by the following of equations:

$$\left. \begin{aligned} \left(N_{L,i}^{\gamma} - N_{L,i}^{(1)}\right)\frac{dx_L}{dt} &= J_i^{\gamma} - J_i^{(1)} \\ \left(N_{R,i}^{(2)} - N_{R,i}^{\gamma}\right)\frac{dx_R}{dt} &= J_i^{(2)} - J_i^{\gamma} \end{aligned} \right\} \tag{13}$$

where: $N_{L,i}^{(1)}, N_{L,i}^{\gamma}, N_{R,i}^{\gamma}, N_{R,i}^{(2)}$ are molar fractions of i-th (i = A, B+C) component on the left (L) and right (R) boundary in Alloy 1, $\gamma$ phase and Alloy 2, respectively; $J_i^{\gamma}$, $J_i^{(1)}$, $J_i^{(2)}$ - diffusion fluxes of i-th component, calculated by the Fick's law, $x_L$ and $x_R$ – location of the left and the right boundary after annealing (fig. 3).

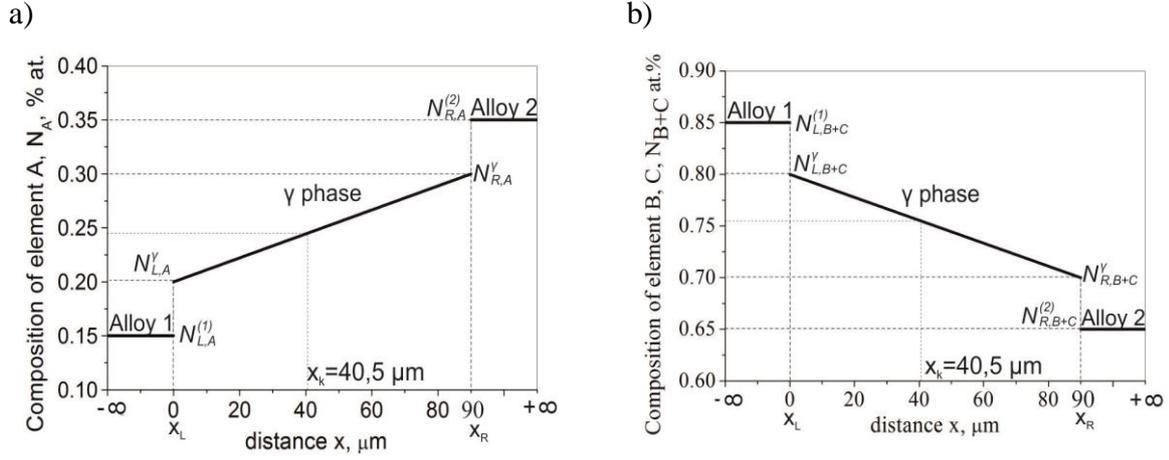

Fig. 3. Composition N of elements: a) A, b) B+C with the location of $x_L$ and $x_R$ boundaries

Assuming further that the fluxes of i-th component in Alloy 1 and Alloy 2 equal to zero: $J_i^{(1)} = J_i^{(2)} = 0$ and $\Delta x = x_R - x_L$, Eq. 9 may be rewritten as:

$$\frac{d\Delta x}{dt} = \frac{-J_i^\gamma}{\left(N_{R,i}^{(2)} - N_{R,i}^\gamma\right)} - \frac{J_i^\gamma}{\left(N_{L,i}^\gamma - N_{L,i}^{(1)}\right)} \tag{14}$$

Because the flux of i-th component can be defined by Fick's law as:

$$J_i^\gamma = -D_i \frac{N_{R,i}^\gamma - N_{L,i}^\gamma}{\Delta x} \tag{15}$$

it is possible to derive the intrinsic diffusion coefficient of i-th component:

$$D_i = \Delta x \frac{d\Delta x}{dt} \left( \frac{\left(N_{R,i}^{(2)} - N_{R,i}^\gamma\right)\left(N_{L,i}^\gamma - N_{L,i}^{(1)}\right)}{\left(N_{R,i}^\gamma - N_{L,i}^\gamma\right)\left(N_{L,i}^\gamma - N_{L,i}^{(1)}\right) + \left(N_{R,i}^\gamma - N_{L,i}^\gamma\right)\left(N_{R,i}^{(2)} - N_{R,i}^\gamma\right)} \right) \tag{16}$$

Considering the data assumed by Paul's, the intrinsic diffusion coefficients (Eq. (16)) have been calculated and the following values were obtained: $D_A = 2.25 \cdot 10^{-14}$ and $D_B = 2.25 \cdot 10^{-14}$ m²/s, respectively. Both values are identical as the differences in concentrations of elements A and B are also equal. Taking this into account, it is obvious that the interdiffusion coefficient calculated in the marker plane (Eq. (6)) yields exactly the same value: $\tilde{D} = 2.25 \cdot 10^{-14}$ m²/s.

In section 2, the interdiffusion and intrinsic diffusion coefficients have been calculated for Paul's data using three independent methods – PBA, AAM, MCA. The values obtained are summarized in Table 2.

Table 2. Calculated values of interdiffusion and intrinsic diffusion coefficients.

|  | Interdiffusion coefficient $\tilde{D}$, m$^2$/s | Intrinsic diffusion coefficient of element A - $D_A$, m$^2$/s | Intrinsic diffusion coefficient of element B - $D_B$, m$^2$/s | Calculated thickness of the intermetallic phase, μm |
|---|---|---|---|---|
| **PBA** | 1.68·10$^{-14}$ | 1.13·10$^{-14}$ | 3.38·10$^{-14}$ | 67.88 |
| **MCA** | 2.25·10$^{-14}$ | 2.25·10$^{-14}$ | 2.25·10$^{-14}$ | 79.76 |
| **AAM** | 2.25·10$^{-14}$ | 1.95·10$^{-14}$ | 3.19·10$^{-14}$ | 81.60 |

In the case of a pseudobinary alloy with only a single intermetallic phase forming, the intermetallic phase growth kinetics can be predicted directly from the mass conservation law at the boundaries, eq. (13). Thus, the following expression should be solved only for one component (e.g. 1):

$$\left. \begin{aligned} \frac{dx_L}{dt} &= \frac{J_i^\gamma}{N_{L,i}^\gamma - N_{L,i}^{(1)}} \\ \frac{dx_R}{dt} &= -\frac{J_i^\gamma}{N_{R,i}^{(2)} - N_{R,i}^\gamma} \end{aligned} \right\} \quad \Rightarrow \quad \frac{dx_R - x_L}{dt} = \frac{-J_i^\gamma \left(N_{L,i}^\gamma - N_{L,i}^{(1)}\right) - J_i^\gamma \left(N_{R,i}^{(2)} - N_{R,i}^\gamma\right)}{\left(N_{L,i}^\gamma - N_{L,i}^{(1)}\right)\left(N_{R,i}^{(2)} - N_{R,i}^\gamma\right)} \quad (17)$$

Table 2 shows also the predicted thickness the intermetallic phase layer formed during the diffusion process. The values were calculated using the diffusion coefficients obtained from different methods.

**2.4. Diffusion coefficients in ternary Ti-NiAl system by MCA method.** In this paragraph the MCA method will be used to approximate the intrinsic diffusion coefficients in ternary multiphase Ti-NiAl system at 1173K. In this system, between pure Ti and NiAl, three intermetallic phases are formed, i.e. AlTi$_3$, Al$_3$NiTi$_2$ and AlNi$_2$Ti [26]. The equilibrium molar fractions of each component in each phase as well as the integral diffusion coefficients and thickness of each layer after annealing for 100 h at 1173 K are shown in figure 4 and summarized in Table 3.

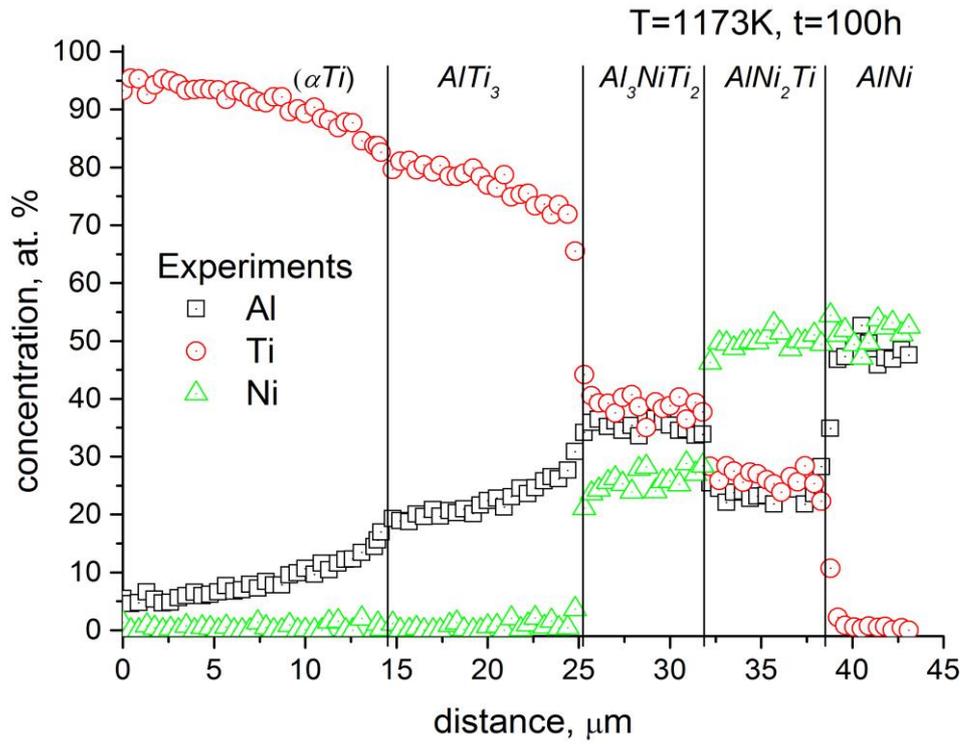

Figure 4. The experimental concentration profile in Ti-NiAl diffusion couple after diffusion at 1173K for 100h.

Table 3. The initial kinetic data –equilibrium molar fractions in Ni-Al-Ti system at 1173K [26]

| Phase | Phase thickness, μm | Equilibrium molar fractions | | |
|---|---|---|---|---|
| | | Ti | Al | Ni |
| $(\alpha Ti)$ | 16,7 | 0.98 - 0.85 | 0.01 - 0.14 | 0.01 – 0.01 |
| $Al\,Ti_3$ | 8,7 | 0.82 - 0.70 | 0.17 - 0.29 | 0.01 – 0.01 |
| $Al_3Ni\,Ti_2$ | 6,8 | 0.44 - 0.36 | 0.33 - 0.35 | 0.23 - 0.29 |
| $Al\,Ni_2Ti$ | 6,2 | 0.28 - 0.22 | 0.22 - 0.28 | 0.5 - 0.5 |
| $Al\,Ni$ | ---- | 0.01 – 0.01 | 0.49 - 0.49 | 0.5 - 0.5 |

The intrinsic diffusion coefficients were predicted by the MCA method. Eq. (13) should be written for each phase boundary and each element. The obtained set of equations is then sloved. In case of the Ti-NiAl system with three different intermetallic phases, the following set of equations should be solved:

$$\left. \begin{aligned} \left( N_{L,i}^{AlTi_3} - N_{R,i}^{Ti} \right) \frac{dx^{Ti|AlTi_3}}{dt} &= J_i^{AlTi_3} - J_i^{Ti} \\ \left( N_{L,i}^{Al_3NiTi_2} - N_{R,i}^{AlTi_3} \right) \frac{dx^{AlTi_3|Al_3NiTi_2}}{dt} &= J_i^{Al_3NiTi_2} - J_i^{AlTi_3} \\ \left( N_{L,i}^{AlNi_2Ti} - N_{R,i}^{Al_3NiTi_2} \right) \frac{dx^{Al_3NiTi_2|AlNi_2Ti}}{dt} &= J_i^{AlNi_2Ti} - J_i^{Al_3NiTi_2} \\ \left( N_{L,i}^{AlNi} - N_{R,i}^{AlNi_2Ti} \right) \frac{dx^{AlNi_2Ti|AlNi}}{dt} &= J_i^{AlNi} - J_i^{AlNi_2Ti} \end{aligned} \right\} \quad (18)$$

where $N_{L,i}^{j}$ and $N_{R,i}^{j}$ are molar fractions of i-th element in j-th phase on the left and right boundary, respectively. The examples of presented notation for concentration of Ti and Al in particular phases of Ti-Ni-Al system are depicted in fig. 5. The fluxes that appear on the right-hand side in the Eq. (18) are calculated by the Fick's law (Eq. (1)) while the thicknesses of the phases used for calculations are given in Table 3. Moreover, as the differences in equilibrium molar fractions in AlNi phase on the left and right boundary of its occurrence equal to 0, also $J_i^{AlNi} = 0$.

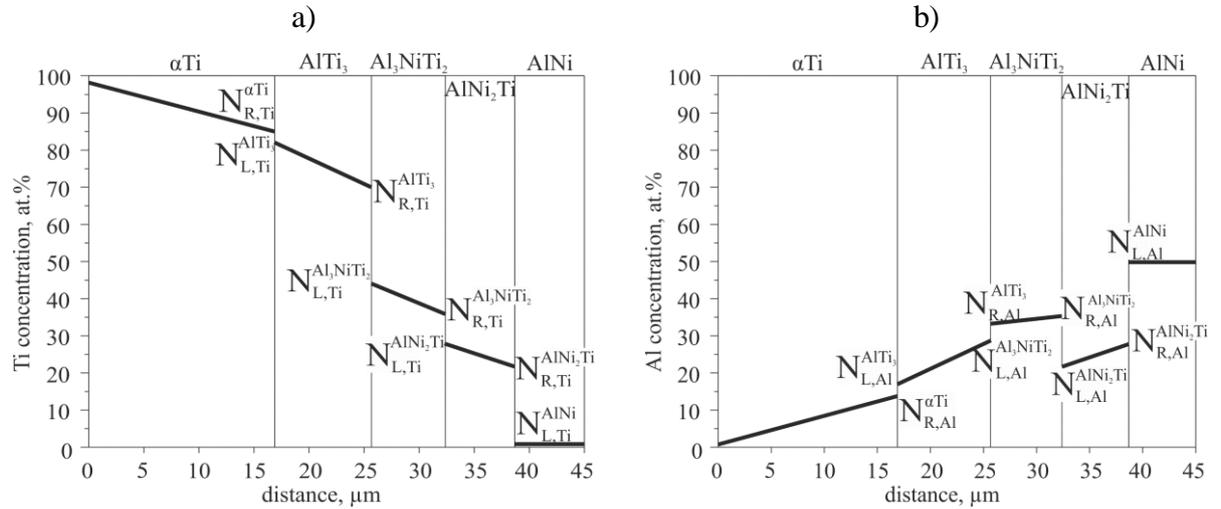

Figure 5. Indication of concentrations of a) Ti and b) Al in particular phases

In the present case it is reasonable to start the solution process from the last, 4$^{th}$ equation, (Eq. (18)). Replacing further by Fick's law (Eq. (1)). The 4$^{th}$ equastion from set of Eq. (18) may be formulated as:

$$\left( N_{L,i}^{AlNi} - N_{R,i}^{AlNi_2Ti} \right) \frac{dx^{AlNi_2Ti|AlNi}}{dt} = D_i^{AlNi_2Ti} \frac{N_{R,i}^{AlNi_2Ti} - N_{L,i}^{AlNi_2Ti}}{\Delta x^{AlNi_2Ti}} \quad (19)$$

The example of calculation of diffusion coefficient of Ti (i = Ti) in AlNi$_2$Ti phase will be presented. The unknown diffusion coefficient may be calculated after reformulation of Eq. 19:

$$D_{Ti}^{AlNi_2Ti} = \frac{\Delta x^{AlNi_2Ti}\left(N_{L,Ti}^{AlNi} - N_{R,Ti}^{AlNi_2Ti}\right)\frac{dx^{AlNi_2Ti|AlNi}}{dt}}{N_{R,Ti}^{AlNi_2Ti} - N_{L,Ti}^{AlNi_2Ti}} \quad (20)$$

The molar fractions of Ti on the left and right boundaries of AlNi and AlNi$_2$Ti phases are as follows: $N_{L,Ti}^{AlNi} = 0.01$, $N_{R,Ti}^{AlNi_2Ti} = 0.22$, $N_{L,Ti}^{AlNi_2Ti} = 0.28$ (table 3, fig. 4).

The thickness of AlNi$_2$Ti phase ($\Delta x^{AlNi_2Ti}$) after diffusion annealing for 100 hours at 1173K, is equal to 6,2 µm (table 3, fig. 4)

The last lacking data, that occurs in Eq. 20 is the displacement of boundary AlNi$_2$Ti|AlNi over the time. Such a derivative has been replaced by the difference between the distance, that this boundary has covered divided by time during the diffusion process. Thus, the derivative from Eq. (20) is calculated as follow: $\frac{dx^{AlNi_2Ti|AlNi}}{dt} = \frac{\Delta x^{AlNi_2Ti}}{t}$ (fig. 5).

According to experimental data (table 3, fig. 4), the displacement of the boundary equals to 6.2 µm at the time $t$ = 100 hours.

Putting together all of the above mentioned data, the diffusion coefficient of Ti in AlNi$_2$Ti can be calculated and it equals: $3.7 \cdot 10^{-12}$ cm$^2 \cdot$s$^{-1}$.

The presented procedure must be then repeated for all other elements, namely Al and Ni. After solution of Eq. (19), the 3$^{rd}$ equation from the set of Eq. 18 may be calculated in the similar way.

The solution of the whole set of Eq. (18) enables to establish intrinsic diffusion coefficients of each element in each phase, except the elements that did not show changes in molar concentrations on particular boundaries. The results of calculations are presented in Table 4.

Table 4. Calculated values of intrinsic diffusion coefficients

| Phase | Intrinsic diffusion coefficients, cm$^2 \cdot$s$^{-1}$ | | |
|---|---|---|---|
| | Ti | Al | Ni |
| ($\alpha$Ti) | $1.7 \cdot 10^{-11}$ | $4.5 \cdot 10^{-12}$ | --- |
| Al Ti$_3$ | $8.3 \cdot 10^{-12}$ | $1.5 \cdot 10^{-12}$ | --- |
| $Al_3Ni\,Ti_2$ | $4.4\cdot10^{-12}$ | $3.9\cdot10^{-12}$ | $4.5\cdot10^{-12}$ |
| $Al\,Ni_2Ti$ | $3.7\cdot10^{-12}$ | $3.7\cdot10^{-12}$ | --- |
| $Al\,Ni$ | --- | --- | --- |

Similarly to the binary system, Eq. (13) has been transformed into Eq. (17) so that the thickness of the growing intermetallic phase could be calculated. Also, the Eq. (18) may be used to calculate the thicknesses of growing phases. In order to solve the inverse problem, Eq. (18) must be reformulated with the assumption that the intrinsic diffusion coefficients in particular phases are known while the phase thicknesses are not. The results of such calculations allow obtaining exactly the same values of thickness of the growing phases as it is demostrated in experimental results (Table 3), which proves the sustainability of the calculations.

**3. Conclusions.** In the present paper, the mass conservation law (MCA method) was used to approximate the intrinsic diffusion coefficients in multiphase ternary system. Currently, there are only few reliable method (even numerical ones, inverse methods) in the literature to approximate the intrinsic diffusion coefficients in multiphase ternary systems. The MCA method was shown to correctly predict the diffusion data even based on a single experimental results. This data can be used as a first approximation of the diffusivities characterizing the process. The MCA method was shown to approximates diffusion coefficients for systems with known difference in chemical composition. In the Ti-NiAl diffusion couple, the diffusion coefficients were established for four phases, i.e. $Al\,Ni_2Ti$, $Al_3Ni\,Ti_2$, $Al\,Ti_3$ and $(\alpha Ti)$. The set of nine diffusion coefficients was determined. The method can be further supplemented by Gibbs energy calculations to overcome the simplifying assumption of chemical activity being equal to the molar fractions.

**Acknowledgment**

This research was financed within the Marie Curie COFUND scheme and POLONEZ program from the National Science Centre, Poland. POLONEZ Grant No. 2015/19/P/ST8/03995. This project has received funding from the European Union's Horizon 2020 research and innovation programme under the Marie Skłodowska-Curie Grant Agreement No. 665778.

**Data Availability.** The data required to reproduce the work reported in the manuscript can be found in Bartek Wierzba - email: bwierzba@prz.edu.pl